\RequirePackage{fixltx2e}
\documentclass[a4paper,fleqn,reqno,english,journal=cmatex,manuscript=article]{revtex4-2}
\usepackage[T1]{fontenc}
\usepackage[latin9]{inputenc}
\usepackage{xcolor}
\usepackage{babel}
\usepackage{array}
\usepackage{float}
\usepackage{textcomp}
\usepackage{multirow}
\usepackage{amsmath}
\usepackage{amsthm}
\usepackage{amssymb}
\usepackage{graphicx}
\usepackage[unicode=true,pdfusetitle,
 bookmarks=true,bookmarksnumbered=false,bookmarksopen=false,
 breaklinks=false,pdfborder={0 0 1},backref=false,colorlinks=false]
 {hyperref}

\makeatletter

\pdfpageheight\paperheight
\pdfpagewidth\paperwidth

\newcommand{\lyxmathsym}[1]{\ifmmode\begingroup\def\b@ld{bold}
  \text{\ifx\math@version\b@ld\bfseries\fi#1}\endgroup\else#1\fi}

\providecommand{\tabularnewline}{\\}

\usepackage{lmodern}

\makeatother

\begin{document}
\title{Probing Optoelectronic Properties of Stable Vacancy-Ordered Double
Perovskites: Insights from Many-Body Perturbation Theory}
\author{Surajit Adhikari{*}, Priya Johari}
\email{sa731@snu.edu.in, priya.johari@snu.edu.in}

\address{Department of Physics, School of Natural Sciences, Shiv Nadar Institution
of Eminence, Greater Noida, Gautam Buddha Nagar, Uttar Pradesh 201314, India.}
\begin{abstract}
A$_{2}$BX$_{6}$ vacancy-ordered double perovskites (VODPs) have
captured substantial research interest in the scientific community
as they offer environmentally friendly and stable alternatives to
lead halide perovskites. In this study, we investigate Rb$_{2}$BCl$_{6}$
(B = Ti, Se, Ru, Pd) VODPs as promising optoelectronic materials employing
state-of-the-art first-principles-based methodologies, specifically
density functional theory combined with density functional perturbation
theory (DFPT) and many-body perturbation theory {[}within the framework
of GW and BSE{]}. Our calculations reveal that all these materials
possess a cubic lattice structure and are both dynamically and mechanically
stable. Interestingly, they all exhibit indirect bandgaps, except
Rb$_{2}$RuCl$_{6}$ displays a metallic character. The G$_{0}$W$_{0}$
bandgap values for these compounds fall within the range of 3.63 to
5.14 eV. Additionally, the results of the BSE indicate that they exhibit
exceptional absorption capabilities across the near-ultraviolet to
mid-ultraviolet light region. Furthermore, studies on transport and
excitonic properties suggest that they exhibit lower effective electron
masses compared to holes, with exciton binding energies spanning between
0.16$-$0.98 eV. We additionally observed a prevalent hole-phonon
coupling compared to electron-phonon coupling in these compounds.
Overall, this study provides valuable insights to guide the design
of vacancy-ordered double perovskites as promising lead-free candidates
for future optoelectronic applications.
\end{abstract}
\maketitle

\section{Introduction:}

Over the past decade, lead halide perovskites APbX$_{3}$ (A = Cs$^{+}$,
CH$_{3}$NH$_{3}$$^{+}$; X = Cl$^{-}$, Br$^{-}$, and I$^{-}$)
have attracted considerable attention for their potential in optoelectronic
applications\citep{chapter1-1,chapter1-2,chapter1-3,chapter1-4}.
Organic-inorganic lead-based hybrid perovskites have reached an impressive
power conversion efficiency (PCE) of up to 26.1\% in perovskite solar
cells\citep{chapter2-12,chapter3-8,chapter3-9,chapter3-10,chapter2-53}.
The rapid increase in PCE (from 3.8\% in 2009 to 26.1\% in 2022) is
attributed to their high absorption coefficient, superior charge-carrier
mobility, defect tolerance, and cost-effective flexible synthesis.
However, their use is constrained by issues such as lead toxicity,
thermal decomposition, and instability when exposed to moisture and
ultraviolet light\citep{chapter3-12,chapter1-16}.

To address this issue, three main stoichiometric classes of perovskites
have garnered substantial research attention in recent years. One
of the classes, specifically double perovskites with the stoichiometry
A$_{2}$BB$'$X$_{6}$ (A = Cs$^{+}$, Rb$^{+}$; B = Ag$^{+}$, Na$^{+}$;
B$'$ = Bi$^{3+}$, In$^{3+}$, Sb$^{3+}$; and X = Cl$^{-}$, Br$^{-}$,
I$^{-}$), is primarily formed through the transmutation of a combination
of monovalent and trivalent metal cations at the B sites\citep{chapter1-19,chapter1-20,chapter1-21}.
For instance, Cs$_{2}$AgInX$_{6}$ (X = Cl, Br)\citep{chapter2-44,chapter1-27},
Cs$_{2}$AgBiX$_{6}$\citep{chapter2-29,chapter2-26}, and Cs$_{2}$AgSbX$_{6}$\citep{chapter2-38,chapter5-18}
are most common examples of double-perovskite materials that have
been widely investigated for a range of optoelectronic applications.
Similarly, there are two other types of structures: A$_{3}$B$_{2}$X$_{9}$
(e.g., Cs$_{3}$Bi$_{2}$I$_{9}$\citep{Conf.5}, Cs$_{3}$Sb$_{2}$I$_{9}$\citep{Conf.6},
etc.) and A$_{2}$BX$_{6}$ (e.g., Cs$_{2}$SnI$_{6}$\citep{Conf.7},
Cs$_{2}$TiI$_{6}$\citep{Conf.8}, etc.). These structures are formed
by substituting trivalent and tetravalent atoms, respectively, with
an empty B site. Here, A$_{2}$BX$_{6}$ is referred to as a vacancy-ordered
double perovskite (VODP), characterized by the removal of every alternating
corner-shared BX$_{6}$ octahedra along all three directions in the
unit cell\citep{Conf.9}.

In recent years, the A$_{2}$BX$_{6}$ VODP family has steadily garnered
considerable attention across various optoelectronic applications
due to their enhanced environmental stability and tunable electronic
and optical properties. For example, Cs$_{2}$BI$_{6}$ (B = Pd, Sn,
Te, Pt) VODPs have been experimentally synthesized as potential photovoltaic
absorbers, exhibiting ideal bandgaps within the range of 1.25 to 1.50
eV\citep{Conf.11,Conf.12,Conf.13,Conf.14}. Furthermore, theoretical
studies indicate that Cs$_{2}$BI$_{6}$ (B = Ti, Pd, Sn, Te, Pt)
compounds exhibit moderate electron mobility ($\sim$1.5$-$51 cm$^{2}$V$^{-1}$s$^{-1}$)
and a high absorption coefficient ($\sim$10$^{5}$ cm$^{-1}$)\citep{Conf.9}.
Consequently, first-principles-based calculations predict a significant
spectroscopic limited maximum efficiency (18$-$23\%), highlighting
their strong potential for applications in optoelectronic devices\citep{Conf.9}.
Similarly, the optoelectronic properties of Rb$_{2}$SnX$_{6}$ VODPs
were also investigated using first-principles calculations and the
bandgaps were found to be 3.52 eV for X = Cl, 2.13 eV for X = Br,
and 1.1 eV for X = I\citep{Conf.10}. Notably, Rb$_{2}$SnI$_{6}$
exhibited exceptional photoelectric properties compared to the other
two compounds, attributed to its optimal bandgap, low effective mass,
and high optical absorption coefficients.

Similarly, Rb$_{2}$BCl$_{6}$ (B = Ti, Se, Pd) VODPs have been successfully
synthesized\citep{Conf.1,Conf.2}. However, theoretical studies on
these compounds are still lacking\citep{Conf.3,Conf.4}. The intrinsic
optoelectronic properties of these VODPs, particularly the roles of
excitonic effects and electron-phonon coupling, are still not fully
understood and remain elusive. The generation of excitons plays a
crucial role in the charge-separation properties of optoelectronic
devices like solar cells. Thus, accurately estimating excitonic properties,
including exciton binding energy, exciton radius, and exciton lifetime,
is essential for these materials. Also, the exciton dynamics and charge
transport are significantly affected by polaronic effects, as carrier
mobility is influenced by the strength of electron-phonon coupling,
which in turn impacts the separation of free charge carriers. Therefore,
understanding the influence of electron-phonon coupling on polaron
mobility is essential, especially in these VODPs.

In this paper, we have therefore performed a systematic and comprehensive
study of Rb$_{2}$BCl$_{6}$ (B = Ti, Se, Ru, Pd) VODPs using state-of-the-art
first-principles-based methodologies. Initially, we assess their stability
and structural properties through density functional theory (DFT)\citep{chapter2-36,chapter2-37}
calculations. Subsequently, we investigate their electronic and optical
properties using many-body perturbation theory (MBPT) within the G$_{0}$W$_{0}$
approximation\citep{chapter1-69,chapter1-70} and the Bethe$-$Salpeter
equation (BSE)\citep{chapter1-67,chapter1-68}, respectively. Further,
the ionic contribution to dielectric function are calculated using
density functional perturbation theory (DFPT)\citep{chapter1-60}.
Finally, we determine the excitonic and polaronic properties using
the Wannier-Mott model\citep{chapter2-38} and the Fr\"ohlich model\citep{chapter2-20},
respectively.

\section{Computational Details:}

In this paper, we conducted state-of-the-art first-principles calculations
utilizing the Vienna Ab initio Simulation Package (VASP)\citep{chapter1-31,chapter1-32}.
Our methodology was grounded in density functional theory (DFT)\citep{chapter2-36,chapter2-37},
density functional perturbation theory (DFPT)\citep{chapter1-60},
and many-body perturbation theory (MBPT)\citep{chapter3-1,chapter3-2}.
In all constituent elements, the interactions of valence electrons
and the atomic core were described using projector-augmented wave
(PAW) pseudopotentials\citep{chapter1-33}. The PAW pseudopotentials
with valence-electron configurations considered for Rb, Ti, Se, Ru,
Pd, and Cl were\textcolor{red}{{} }4s$^{2}$4p$^{6}$5s$^{1}$, 3p$^{6}$4s$^{2}$3d$^{2}$,
4s$^{2}$4p$^{4}$, 4p$^{6}$5s$^{2}$4d$^{6}$, 5s$^{2}$4d$^{8}$,\textcolor{red}{{}
}and 3s$^{2}$3p$^{5}$, respectively. For the structural optimization,
the exchange-correlation (xc) functional of Perdew, Burke, and Ernzerhof
(PBE) based on the generalized gradient approximation (GGA)\citep{chapter1-34}
was employed to account for electron-electron interactions. The plane-wave
cutoff energy was configured to 400 eV, and the convergence threshold
for the electronic self-consistent field iterations was set at 10$^{-6}$
eV. The lattice constants and atomic coordinates were fully optimized
until the Hellmann-Feynman forces on each atom were reduced to less
than 0.01 eV/\AA. The crystal structures were optimized using a $\Gamma$-centered
$4\times4\times4$ $\mathbf{k}$-point sampling scheme for Brillouin
zone integration. The optimized structures were then visualized using
the Visualization for Electronic and STructural Analysis (VESTA) software
package\citep{chapter2-3}. The phonon spectra were computed employing
the DFPT method utilizing $2\times2\times2$ supercells, as implemented
in the PHONOPY package\citep{chapter3-6}.\textcolor{purple}{{} }The
electronic band structures were analyzed through computations employing
the hybrid HSE06\citep{chapter1-35} xc functional and the many-body
perturbation theory (MBPT) based GW (G$_{0}$W$_{0}$@PBE) method\citep{chapter1-69,chapter1-70}.
It's worth mentioning that the spin-orbit coupling (SOC) effect was
omitted, as it doesn't influence the bandgap. The effective mass was
determined by SUMO\citep{chapter2-10}, utilizing a parabolic fitting
approach on the band edges.\textcolor{purple}{{} }Additionally, we conducted
Bethe-Salpeter equation (BSE)\citep{chapter1-67,chapter1-68} calculations,
building upon the single-shot GW (G$_{0}$W$_{0}$)@PBE method to
accurately gauge optical properties, which explicitly addressing electron-hole
interactions. For these GW-BSE calculations, a $4\times4\times4$
$\Gamma$-centered $\mathbf{k}$-grid and a converged NBANDS value
of 540 were employed. The electron-hole kernel for BSE computations
was constructed by considering 6 occupied and 6 unoccupied bands.\textcolor{purple}{{}
}The elastic and optical properties were post-processed using the
VASPKIT\citep{chapter1-48} package, and the DFPT method was employed
to calculate the ionic contribution to the dielectric constant.

\section{Results and Discussions:}

In this study, we conducted a systematic and comprehensive investigation
of Rb$_{2}$BCl$_{6}$ (B = Ti, Se, Ru, Pd) vacancy-ordered double
perovskites to examine their potential optoelectronic properties.
In the following subsections, we explore and discuss in detail the
stability as well as the structural, electronic, optical, excitonic,
and polaronic properties of Rb$_{2}$BCl$_{6}$ VODPs to provide a
fundamental understanding and guide future experimental studies.

\begin{figure}[H]
\begin{centering}
\includegraphics[width=1\textwidth,height=1\textheight,keepaspectratio]{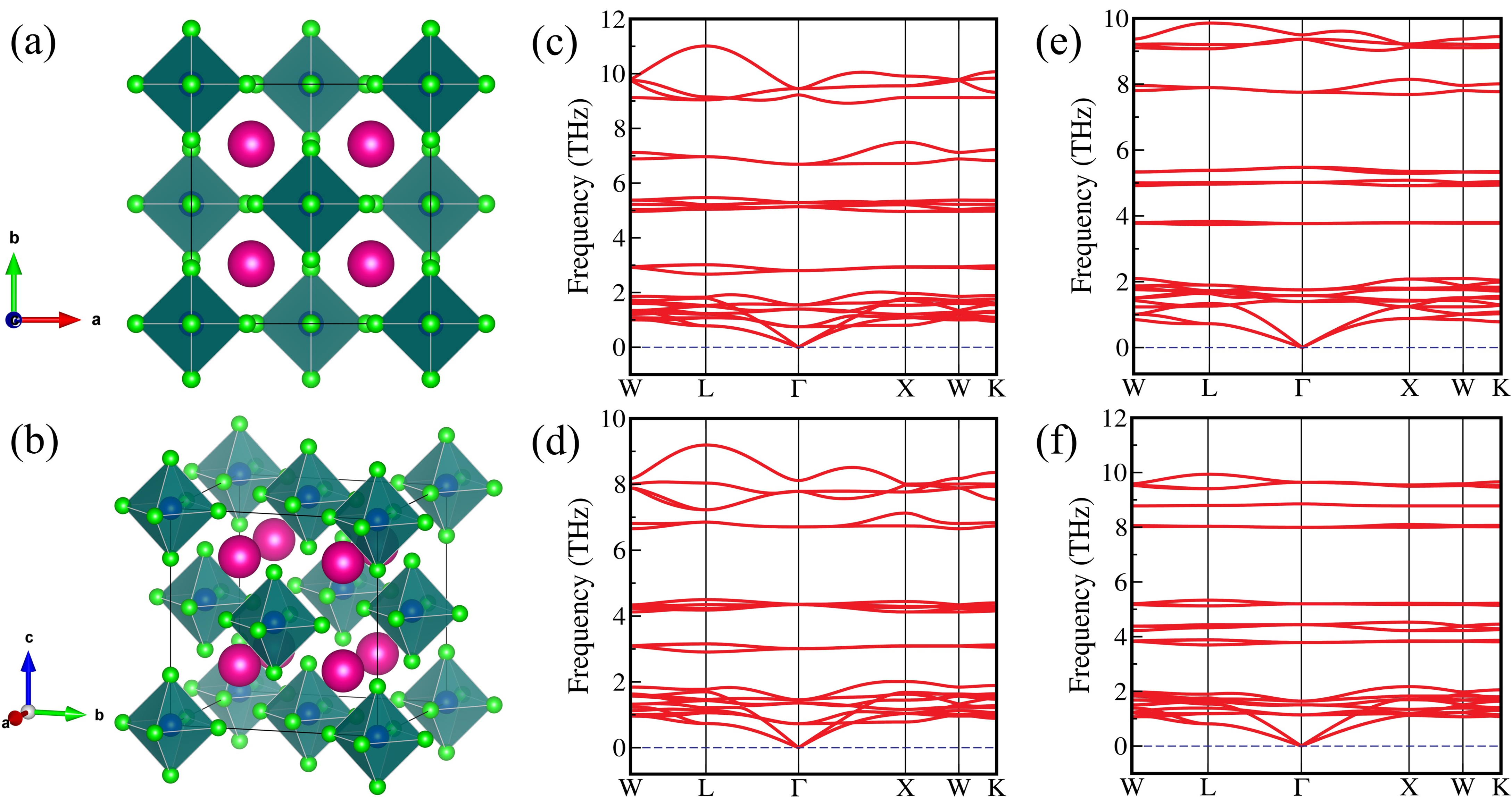}
\par\end{centering}
\caption{\label{fig:1}(a, b) Crystal structure of Rb$_{2}$BCl$_{6}$ (B =
Ti, Se, Ru, Pd) VODPs in face-centered cubic phase, and phonon dispersion
curves of (c) Rb$_{2}$TiCl$_{6}$, (d) Rb$_{2}$SeCl$_{6}$, (e)
Rb$_{2}$RuCl$_{6}$, and (f) Rb$_{2}$PdCl$_{6}$ calculated with
the DFPT method. Pink, blue, and green balls represent Rb, Ti/Se/Ru/Pd,
and Cl atoms, respectively.}
\end{figure}

\subsection{Structural Properties:}

Rb$_{2}$BCl$_{6}$ (B = Ti, Se, Ru, Pd) VODPs possess a face-centered
cubic crystal structure with the space group Fm$\bar{3}$m (No. 225).
The geometric configuration of Rb$_{2}$BCl$_{6}$ is depicted in
Figure \ref{fig:1}. In this structure, the Rb atoms are located at
the 8c Wyckoff positions with coordinates (0.25, 0.25, 0.25), the
B atoms are positioned at the 4a Wyckoff positions with coordinates
(0, 0, 0), and the Cl atoms are situated at the 24e Wyckoff positions
with coordinates (x, 0, 0), where the value of x is approximately
0.20. Each Rb atom is positioned between the {[}BCl$_{6}${]} octahedra,
surrounded by twelve Cl atoms. Meanwhile, the B atoms occupy the corners
and face-centered positions of the {[}BCl$_{6}${]} octahedra. The
optimized lattice parameters and bond lengths of Rb$_{2}$BCl$_{6}$
VODPs are presented in Table \ref{tab:1}, and it is found that lattice
parameters exhibit good agreement with experiments.

First, we have assessed the stability of the material, as it is a
crucial factor for achieving high-performance device applications.
We have calculated the Goldschmidt tolerance factor ($t$) and the
octahedral factor ($\mu$) to predict the structural stability of
these VODPs\citep{chapter1-36,chapter1-37,chapter1-38}. Recently,
Bartel et al. proposed a new tolerance factor\citep{chapter1-39}
($\tau$), which is also computed for these VODPs. These three parameters
are defined as follows:

\begin{equation}
t=\frac{r_{A}+r_{X}}{\sqrt{2}(r_{B}+r_{X})},\mu=\frac{r_{B}}{r_{X}},
\end{equation}

\begin{center}
\begin{equation}
\tau=\frac{r_{X}}{r_{B}}-n_{A}\left(n_{A}-\frac{r_{A}/r_{B}}{ln(r_{A}/r_{B})}\right),
\end{equation}
\par\end{center}

where $r_{A}$, $r_{B}$, and $r_{X}$ are the Shannon ionic radii
for A (Rb), B (Ti, Se, Ru, Pd), and X (Cl) ions, while $n_{A}$ denotes
the oxidation state of the A (Rb)-site cation. For a stable perovskite
structure, the values should fall within the following ranges: $0.442<\mu<0.895$,
$0.80<t<1.10$, and $\tau<4.18$\citep{chapter1-37,chapter2-2,chapter1-39}.
Since the values of $\mu$ and $\tau$ are primarily applicable to
ABX$_{3}$ and A$_{2}$BB'X$_{6}$ compounds, deviations may occur
for VODPs due to the presence of defects. The calculated values indicate
that these VODPs are stable in their cubic structures (see Table \ref{tab:1}).

\begin{table}[H]
\caption{\label{tab:1}Optimized lattice parameter (in $\textrm{\AA}$), bond
lengths (in $\textrm{\AA}$), Goldschmidt tolerance factor ($t$), octahedral
factor ($\mu$), and new tolerance factor ($\tau$) of Rb$_{2}$BCl$_{6}$
(B = Ti, Se, Ru, Pd) VODPs are listed.}

\centering{}{\footnotesize{}}%
\begin{tabular}{cccccccccc}
\hline 
\multirow{2}{*}{{\footnotesize{}Compounds}} & \multicolumn{2}{c}{{\footnotesize{}Lattice parameter (a = b = c), $\textrm{\AA}$}} & \multirow{2}{*}{} & \multicolumn{2}{c}{{\footnotesize{}Bond lengths, $\textrm{\AA}$}} & \multirow{2}{*}{} & \multirow{2}{*}{{\footnotesize{}$t$}} & \multirow{2}{*}{{\footnotesize{}$\mu$}} & \multirow{2}{*}{{\footnotesize{}$\tau$}}\tabularnewline
\cline{2-3} \cline{3-3} \cline{5-6} \cline{6-6} 
 & {\footnotesize{}In our study} & {\footnotesize{}Experimental report} &  & {\footnotesize{}Rb-Cl} & {\footnotesize{}B-Cl} &  &  &  & \tabularnewline
\hline 
{\footnotesize{}Rb$_{2}$TiCl$_{6}$} & {\footnotesize{}10.24} & {\footnotesize{}9.92\citep{Conf.2}} &  & {\footnotesize{}3.62} & {\footnotesize{}2.36} &  & {\footnotesize{}1.034} & {\footnotesize{}0.334} & {\footnotesize{}4.434}\tabularnewline
{\footnotesize{}Rb$_{2}$SeCl$_{6}$} & {\footnotesize{}10.32} & {\footnotesize{}9.98\citep{Conf.2}} &  & {\footnotesize{}3.65} & {\footnotesize{}2.43} &  & {\footnotesize{}1.081} & {\footnotesize{}0.276} & {\footnotesize{}5.189}\tabularnewline
{\footnotesize{}Rb$_{2}$RuCl$_{6}$} & {\footnotesize{}10.09} & {\footnotesize{}\textendash{}} &  & {\footnotesize{}3.57} & {\footnotesize{}2.34} &  & {\footnotesize{}1.027} & {\footnotesize{}0.343} & {\footnotesize{}4.357}\tabularnewline
{\footnotesize{}Rb$_{2}$PdCl$_{6}$} & {\footnotesize{}10.19} & {\footnotesize{}9.99\citep{Conf.1}} &  & {\footnotesize{}3.61} & {\footnotesize{}2.34} &  & {\footnotesize{}1.029} & {\footnotesize{}0.340} & {\footnotesize{}4.382}\tabularnewline
\hline 
\end{tabular}{\footnotesize\par}
\end{table}

In addition to assessing structural stability, we have also evaluated
the dynamical and mechanical stabilities. We have calculated self-consistent
phonon calculations for these VODPs for dynamical stability using
the DFPT method\citep{chapter1-60}. Dynamical stability is a key
indicator of a material's stability as it is associated with the phonon
modes. For VODPs, the structural symmetry reveals 27 phonon modes
due to 9 atoms per unit cell. Out of the 27 phonon modes, 3 are acoustic,
while the rest are optical modes, categorized into low- and high-frequency
phonons. The absence of imaginary frequencies confirms the dynamical
stability of these Rb$_{2}$BCl$_{6}$ perovskites at T = 0 K {[}see
Figure \ref{fig:1}(c)-(f){]}.

\begin{table}[H]
\caption{\label{tab:2}Elastic parameters for Rb$_{2}$BCl$_{6}$ (B = Ti,
Se, Ru, Pd) VODPs, including elastic constants $C_{ij}$ (in GPa),
bulk modulus $B$ (in GPa), shear modulus $G$ (in GPa), Young\textquoteright s
modulus $Y$ (in GPa), Poisson\textquoteright s ratio $\nu$ (dimensionless),
and Zener anisotropic factor $A$ (dimensionless).}

\centering{}%
\begin{tabular}{cccccccccc}
\hline 
Compounds & $C_{11}$ & $C_{12}$ & $C_{44}$ & $B$ & $G$ & $Y$ & $B/G$ & $\nu$ & $A$\tabularnewline
\hline 
Rb$_{2}$TiCl$_{6}$ & 6.68 & 1.03 & 5.45 & 2.91 & 4.19 & 8.49 & 0.70 & 0.01 & 1.93\tabularnewline
Rb$_{2}$SeCl$_{6}$ & 19.91 & 2.96 & 5.50 & 8.61 & 6.54 & 15.66 & 1.32 & 0.20 & 0.65\tabularnewline
Rb$_{2}$RuCl$_{6}$ & 19.52 & 4.66 & 6.29 & 9.61 & 6.72 & 16.35 & 1.43 & 0.22 & 0.85\tabularnewline
Rb$_{2}$PdCl$_{6}$ & 28.28 & 17.37 & 7.04 & 21.01 & 6.36 & 17.32 & 3.31 & 0.36 & 1.29\tabularnewline
\hline 
\end{tabular}
\end{table}

Subsequently, we calculated the elastic constants ($C_{ij}$) of these
perovskites to determine their mechanical stability using the energy-strain
approach\citep{chapter1-47}. Since all VODPs considered in this study
are cubic, three independent elastic constants, such as $C_{11}$,
$C_{12}$, and $C_{44}$, are sufficient to explain the mechanical
stability and related properties of the material. According to the
Born stability criteria\citep{chapter1-47}, the necessary and sufficient
conditions for the mechanical stability of cubic crystal systems are
as follows:

\begin{equation}
C_{11}>0,C_{11}-C_{12}>0,C_{11}+2C_{12}>0,C_{44}>0
\end{equation}

The calculated $C_{ij}$ values for all VODPs are listed in Table
\ref{tab:2}, and they satisfy the Born stability conditions, indicating
excellent mechanical stability in the cubic phase. Using these elastic
constants, we calculated the bulk modulus ($B$), shear modulus ($G$),
and Young\textquoteright s modulus ($Y$) of the perovskites based
on the Voigt\textminus Reuss\textminus Hill approaches\citep{chapter1-49,chapter1-50}
(see Table \ref{tab:2}). Additionally, Pugh\textquoteright s ratio
($B/G$) and Poisson\textquoteright s ratio ($\nu$) are utilized
to examine the fragility of these perovskites. Materials exhibit ductile
behavior when the $B/G$ and $\nu$ values exceed 1.75 and 0.26, respectively;
otherwise, they are considered brittle. From Table \ref{tab:2}, it
is clear that all examined VODPs are brittle, except Rb$_{2}$PdCl$_{6}$.
We also evaluated the Zener anisotropic factor ($A$) of these materials
using the relation: $A=2C_{44}/(C_{11}-C_{12})$. The value of $A$
is 1 for an isotropic system. Deviations from this value indicate
the degree of elastic anisotropy in the crystal. Based on the calculated
values, all the examined VODPs exhibit anisotropic behavior.

\subsection{Electronic Properties:}

Once stability is confirmed, electronic structure calculations for
Rb$_{2}$BCl$_{6}$ VODPs are also performed, as these are essential
for the design of optoelectronic devices. Here, we estimate the partial
density of states (PDOS), total density of states (TDOS), the positions
of the band edges$\lyxmathsym{\textemdash}$specifically, the conduction
band minima (CBM) and valence band maxima (VBM), as well as the nature
of the bandgap to gain a deeper understanding of the electronic structure.

\begin{figure}[H]
\begin{centering}
\includegraphics[width=1\textwidth,height=1\textheight,keepaspectratio]{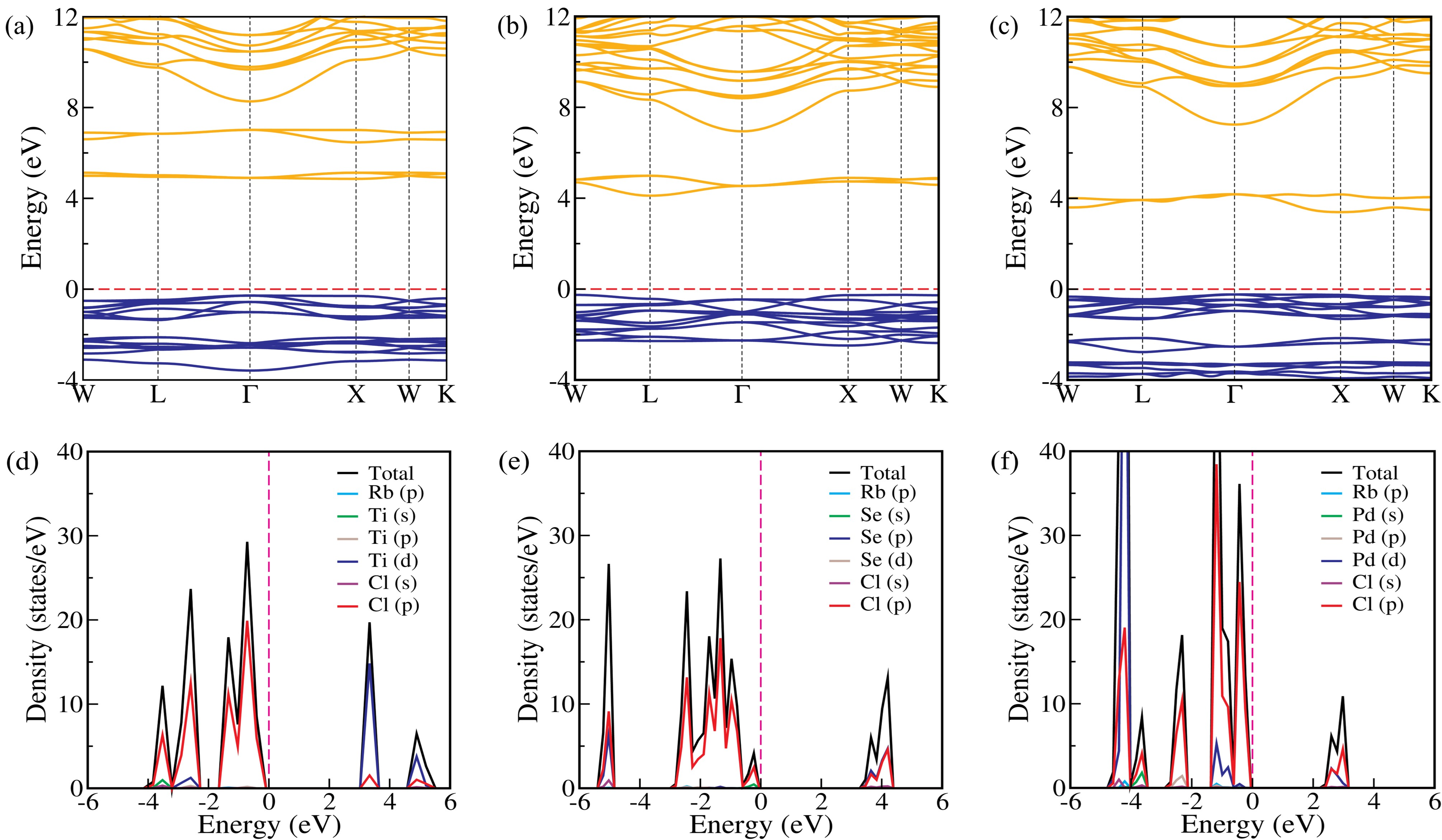}
\par\end{centering}
\caption{\label{fig:2}Calculated band structures of (a) Rb$_{2}$TiCl$_{6}$,
(b) Rb$_{2}$SeCl$_{6}$, and (c) Rb$_{2}$PdCl$_{6}$ VODPs obtained
using G$_{0}$W$_{0}$@PBE method, as well as TDOS and PDOS of (d)
Rb$_{2}$TiCl$_{6}$, (e) Rb$_{2}$SeCl$_{6}$, and (f) Rb$_{2}$PdCl$_{6}$
VODPs obtained using HSE06 xc functional, respectively. The Fermi
level is set to be zero and marked by the dashed line.}
\end{figure}

Initially, electronic structure calculations for Rb$_{2}$BCl$_{6}$
VODPs are conducted using the widely adopted semilocal PBE xc functional,
both with and without accounting for spin-orbit coupling (SOC) effects.
However, it is well established that PBE underestimates the bandgap
of double perovskites due to the self-interaction error\citep{chapter2-38,chapter5-18},
and SOC does not alter bandgap significantly (see Table \ref{tab:3}).
Therefore, we also employed the hybrid HSE06 xc functional\citep{chapter1-35}
and the many-body perturbation theory (MBPT) based GW method, specifically
the G$_{0}$W$_{0}$@PBE approach\citep{chapter1-69,chapter1-70}
to achieve a more accurate estimation of bandgaps. The G$_{0}$W$_{0}$@PBE
band structure of the primitive cell of Rb$_{2}$BCl$_{6}$ (B = Ti,
Se, Pd) VODPs are depicted in Figure \ref{fig:2}(a)-(c), while Rb$_{2}$RuCl$_{6}$
exhibits metallic character. The VBMs and CBMs of Rb$_{2}$BCl$_{6}$
VODPs are localized at different k-points within the Brillouin zone,
resulting in these materials having an indirect band gap. The bandgap
values of these compounds, calculated using the PBE/PBE-SOC, HSE06
xc functionals, and the G$_{0}$W$_{0}$@PBE method, are presented
in Table \ref{tab:3}. Based on this table, it is evident that our
HSE06 bandgap values are in good agreement with previous theoretical
results\citep{Conf.3,Conf.4}. Also, we computed the value of the
lowest direct bandgap ($E_{g}^{dir}$) of these compounds to estimate
the exciton binding energy, as detailed in Table \ref{tab:4}.

Figure \ref{fig:2}(a) depicts the electronic G$_{0}$W$_{0}$@PBE
band structure of Rb$_{2}$TiCl$_{6}$ VODP, which reveals an indirect
bandgap of 5.14 eV, with the CBM and VBM located at the X and $\Gamma$
points, respectively, within the first Brillouin zone. The PDOS calculation
of this compound shows that the VBM is predominantly composed of Cl-3p
orbital, while the CBM is largely influenced by Ti-3d orbital with
a minor contribution of Cl-3p orbital {[}Figure \ref{fig:2}(d){]}.
Subsequently, Rb$_{2}$SeCl$_{6}$ exhibits an indirect G$_{0}$W$_{0}$@PBE
bandgap of 4.36 eV, and the CBM at the L and VBM at the W points {[}Figure
\ref{fig:2}(b){]}. Here, the CBM is mainly composed of Se-4p and
Cl-3p orbitals, whereas the VBM mainly originates from the Cl-3p state,
with a minor amount of Se-4s orbital {[}Figure \ref{fig:2}(e){]}.
Ongoing from Se to Pd for B in Rb$_{2}$BCl$_{6}$, the bandgap remains
indirect, but its value decreases to 3.63 eV, and the positions of
VBM and CBM remain the same as in Rb$_{2}$TiCl$_{6}$ {[}Figure \ref{fig:2}(c){]}.
Similar to Ti-based VODP, the VBM of Pd-based VODP is primarily composed
of Cl-3p orbital, and the CBM is mainly formed by the antibonding
hybridized states of Pd-4d and Cl-3p orbitals {[}Figure \ref{fig:2}(f){]}.
Overall, these materials exhibit a wide range of bandgaps, making
them promising candidates for optoelectronic devices.

\begin{table}[H]
\caption{\label{tab:3}Bandgap (in eV) calculated using the PBE, HSE06, and
G$_{0}$W$_{0}$@PBE method, respectively, as well as computed effective
mass of electron ($m_{e}^{*}$) and hole ($m_{h}^{*}$) and reduced
mass ($\mu^{*}$) of Rb$_{2}$BCl$_{6}$ (B = Ti, Se, Pd) VODPs. Here,
all values of the effective mass are in terms of free-electron mass
($m_{0}$).}

\centering{}{\scriptsize{}}%
\begin{tabular}{cccccccc}
\hline 
{\scriptsize{}Compounds} & {\scriptsize{}PBE/PBE-SOC} & {\scriptsize{}HSE06} & {\scriptsize{}G$_{0}$W$_{0}$@PBE} & {\scriptsize{}Previous work} & {\scriptsize{}$m_{e}^{*}$ ($m_{0}$)} & {\scriptsize{}$m_{h}^{*}$ ($m_{0}$)} & {\scriptsize{}$\mu^{*}$ ($m_{0}$)}\tabularnewline
\hline 
{\scriptsize{}Rb$_{2}$TiCl$_{6}$} & {\scriptsize{}2.17/2.14} & {\scriptsize{}3.54} & {\scriptsize{}5.14} & {\scriptsize{}3.61\citep{Conf.3}, 3.07\citep{Conf.4}} & {\scriptsize{}2.007 (}\textbf{\scriptsize{}2.007}{\scriptsize{})} & {\scriptsize{}2.942 (}\textbf{\scriptsize{}1.505}{\scriptsize{})} & {\scriptsize{}1.193 (}\textbf{\scriptsize{}0.860}{\scriptsize{})}\tabularnewline
{\scriptsize{}Rb$_{2}$SeCl$_{6}$} & {\scriptsize{}2.61/2.58} & {\scriptsize{}3.39} & {\scriptsize{}4.36} & {\scriptsize{}3.07\citep{Conf.3}, 3.16\citep{Conf.4}} & {\scriptsize{}0.755 (}\textbf{\scriptsize{}0.755}{\scriptsize{})} & {\scriptsize{}3.168 (}\textbf{\scriptsize{}1.200}{\scriptsize{})} & {\scriptsize{}0.610 (}\textbf{\scriptsize{}0.463}{\scriptsize{})}\tabularnewline
{\scriptsize{}Rb$_{2}$PdCl$_{6}$} & {\scriptsize{}1.33/1.25} & {\scriptsize{}2.43} & {\scriptsize{}3.63} & {\scriptsize{}2.38\citep{Conf.3}} & {\scriptsize{}0.849 (}\textbf{\scriptsize{}0.849}{\scriptsize{})} & {\scriptsize{}2.972 (}\textbf{\scriptsize{}1.505}{\scriptsize{})} & {\scriptsize{}0.660 (}\textbf{\scriptsize{}0.543}{\scriptsize{})}\tabularnewline
\hline 
\end{tabular}{\scriptsize\par}
\end{table}

To gain deeper insight into charge carrier transport, we also computed
the effective masses of electrons ($m_{e}^{*}$) and holes ($m_{h}^{*}$)
for all the VODPs. The effective masses are estimated by fitting the
E\textminus k dispersion curves derived from the G$_{0}$W$_{0}$@PBE
band structure calculations using the formula, $\text{\ensuremath{m^{*}=\hbar^{2}\left[\partial^{2}E(k)/\partial k^{2}\right]^{-1}}}$and
are tabulated in Table \ref{tab:3}. The data clearly shows that the
electron effective masses for these VODPs are smaller than those of
the holes, suggesting higher electron mobility compared to hole mobility.

\subsection{Optical Properties:}

Apart from the electronic properties, we also compute the optical
response of the VODPs, which is a crucial attribute for the perovskite
optoelectronic devices. To achieve higher reliability in our estimations,
we performed Bethe-Salpeter equation (BSE) simulations\citep{chapter1-67,chapter1-68},
based on the single-shot GW (G$_{0}$W$_{0}$)@PBE method, which explicitly
incorporates for electron-hole interactions. The optical response
of the Rb$_{2}$BCl$_{6}$ VODPs is evaluated by calculating the frequency-dependent
($\omega$) dielectric function, $\varepsilon$($\omega$) = {[}Re($\varepsilon$){]}
+ i{[}Im($\varepsilon$){]}, where {[}Re($\varepsilon$){]} and {[}Im($\varepsilon$){]}
represent the real and imaginary parts of dielectric function, respectively.

The real part of the dielectric function, {[}Re($\varepsilon$){]},
describes how a material polarizes in response to an electric field.
High values of {[}Re($\varepsilon$){]} signify strong polarization,
which can influence the material's optical properties, such as its
refractive index and light absorption. Additionally, {[}Re($\varepsilon$){]}
at zero energy, often referred to as the electronic (or optical) dielectric
constant ($\varepsilon_{\infty}$), characterizes dielectric screening
during Coulomb interactions between electrons and holes. The values
of $\varepsilon_{\infty}$ for the materials considered, calculated
using the BSE@G$_{0}$W$_{0}$@PBE method, are tabulated in Table
\ref{tab:4}. A high static dielectric constant is associated with
a lower rate of charge carrier recombination, which can enhance the
efficiency of optoelectronic devices\citep{chapter2-48}.

\begin{figure}[H]
\begin{centering}
\includegraphics[width=1\textwidth,height=1\textheight,keepaspectratio]{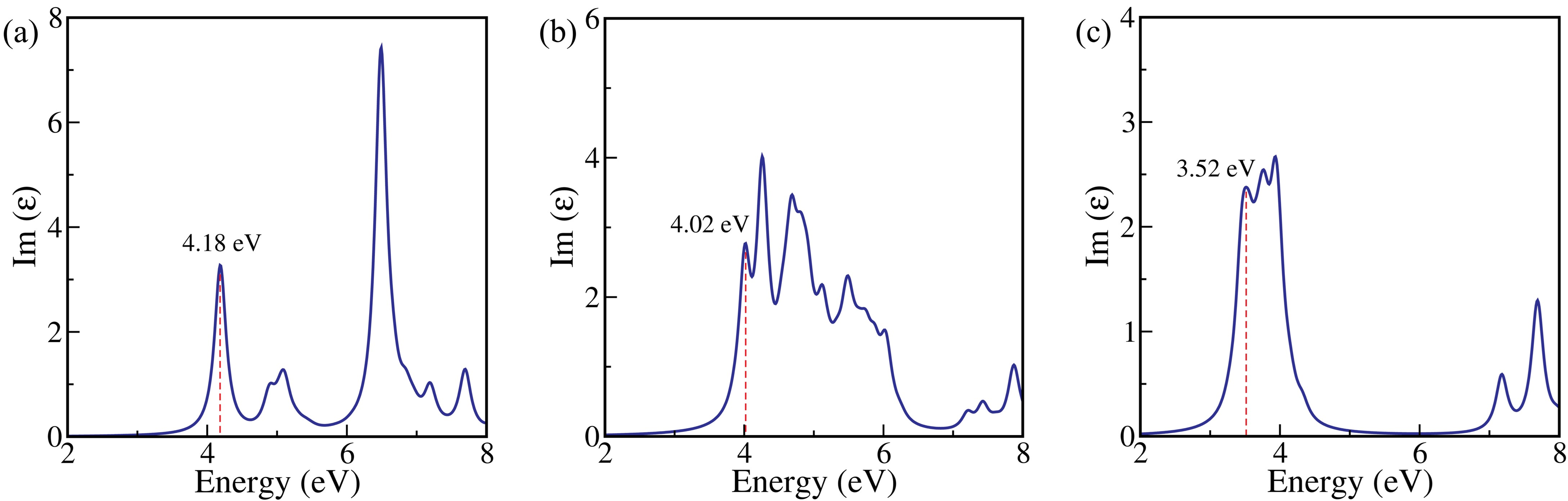}
\par\end{centering}
\caption{\label{fig:3}Imaginary part {[}Im($\varepsilon$){]} of the dielectric
function for (a) Rb$_{2}$TiCl$_{6}$, (b) Rb$_{2}$SeCl$_{6}$, and
(c) Rb$_{2}$PdCl$_{6}$ VODPs obtained using BSE@G$_{0}$W$_{0}$@PBE
method.}
\end{figure}

The imaginary part of the dielectric function, {[}Im($\varepsilon$){]},
is crucial for predicting the linear absorption properties of a material.
Figure \ref{fig:3} shows the imaginary part of the dielectric function
for Rb$_{2}$BCl$_{6}$ VODPs, as obtained using the BSE@G$_{0}$W$_{0}$@PBE
method. The intercept along the energy axis in a plot of the imaginary
part of the frequency-dependent dielectric function versus energy
can be used to determine the absorption onset (edge) of VODPs. It
is observed that these compounds exhibit an absorption onset in the
range of $\sim$ 3.30$-$3.90 eV, indicating ultraviolet light absorption
(see Figure \ref{fig:3}). Overall, optical properties suggest that
Rb$_{2}$BCl$_{6}$ VODPs could be promising candidates for the UV
optoelectronic applications including UV sensors, UV LEDs, and photodetectors.

\subsection{Excitonic Properties:}

In addition, we examined the exciton properties of these VODPs by
calculating the exciton binding energy, exciton radius, and exciton
lifetime to assess their potential for optoelectronic applications.
The exciton binding energy is defined as the energy required to separate
an exciton into its constituent electron (e) and hole (h) pair.\textcolor{red}{{}
}The exciton binding energy ($E_{B}$) for a screened Coulomb interacting
$e-h$ pair is calculated using the hydrogenic Wannier\textminus Mott
(WM)\citep{chapter2-38} model, as follows:

\begin{equation}
\mathrm{E_{B}=\left(\frac{\mu^{*}}{m_{0}\varepsilon_{eff}^{2}}\right)R_{\infty}},\label{eq:1}
\end{equation}

where, $\mu^{*}$ represents the reduced mass of the charge carriers
at the direct band edges, $m_{0}$ denotes the rest mass of an electron,
$\mathrm{\varepsilon_{eff}}$ is the effective dielectric constant,
and $R_{\infty}$ is the Rydberg constant (13.6 eV). The reduced mass
of the carrier ($\mu^{*}$) is given by,

\begin{equation}
\mathrm{\frac{1}{\mu^{*}}=\frac{1}{m_{e}^{*}}+\frac{1}{m_{h}^{*}}}\label{eq:20}
\end{equation}

To determine the $E_{B}$ for the VODPs under study, we need to calculate
the effective dielectric constant ($\mathrm{\varepsilon_{eff}}$).
Here, $\mathrm{\varepsilon_{eff}}$ is situated between the electronic
(optical) dielectric constant ($\varepsilon_{\infty}$) and the static
(electronic + ionic) dielectric constant ($\varepsilon_{static}$).
The upper and lower bounds of the $E_{B}$ are determined by the electronic
and static dielectric constants, respectively. Note that, $\varepsilon_{\infty}$
for these compounds is calculated using BSE method, while the ionic
part of the dielectric constants is determined using DFPT method\citep{chapter1-60}.
Using the preceding equation \ref{eq:1}, we have computed the upper
bound ($E_{Bu}$) and the lower bound ($E_{Bl}$) of the exciton binding
energy, which are tabulated in Table \ref{tab:4}.

Besides the WM model, we further computed the $E_{B}$ from standard
BSE calculations. Theoretically, $E_{B}$ is computed as the difference
between the energy of the unbounded noninteracting e-h pair (i.e.,
direct G$_{0}$W$_{0}$ bandgap) and the energy of the bound e-h pair
(i.e., the BSE peak position)\citep{chapter5-16,chapter5-18}. In
the BSE method, the e-h pair is bound by the screened Coulomb interaction,
and the energy of the lowest peak position is referred to as the exciton
energy ($E_{o}$). The $E_{B}$ values for these VODPs, obtained from
standard BSE calculations, are tabulated in Table \ref{tab:4}. From
Tables \ref{tab:4} and \ref{tab:5}, it is evident that $E_{B}\gg\hbar\omega_{LO}$,
where $\omega_{LO}$ denotes the characteristic phonon angular frequency.
Thus, the ionic contribution to the static dielectric constant is
negligible, leading to $\mathrm{\varepsilon_{eff}}\rightarrow\varepsilon_{\infty}$,
which implies that the static dielectric constant mainly consists
of electronic contribution at a high frequency ($\varepsilon_{\infty}$)\citep{chapter1-65,chapter1-66}.

\begin{table}[H]
\caption{\label{tab:4}Calculated excitonic parameters of Rb$_{2}$BCl$_{6}$
(B = Ti, Se, Pd) VODPs. Here, $\varepsilon_{\infty}$ and $\varepsilon_{static}$
represents the electronic and the static dielectric constant, $E_{Bu}$
and $E_{Bl}$ represents the upper and lower bound of exciton binding
energy, $r_{exc}$ represents exciton radius, $|\phi_{n}(0)|^{2}$
represents the probability of a wave function for the electron-hole
pair at zero charge separation, $E_{g}^{dir}$ represents\textbf{
}the direct G$_{0}$W$_{0}$@PBE bandgap, $E_{o}$ represents the
exciton energy, $E_{B}$ represents the exciton binding energy, and
$T_{exc}$ represents the excitonic temperature.}

\centering{}{\tiny{}}%
\begin{tabular}{cccccccccccc}
\hline 
\multirow{2}{*}{{\tiny{}Compounds}} & \multicolumn{6}{c}{{\tiny{}Wannier-Mott Model}} & \multirow{2}{*}{} & \multicolumn{4}{c}{{\tiny{}Standard BSE}}\tabularnewline
\cline{2-7} \cline{3-7} \cline{4-7} \cline{5-7} \cline{6-7} \cline{7-7} \cline{9-12} \cline{10-12} \cline{11-12} \cline{12-12} 
 & {\tiny{}$\varepsilon_{\infty}$} & {\tiny{}$E_{Bu}$ (eV)} & {\tiny{}$\varepsilon_{static}$} & {\tiny{}$E_{Bl}$ (eV)} & {\tiny{}$r_{exc}$ (nm)} & {\tiny{}$|\phi_{n}(0)|^{2}$ (10$^{28}$ m$^{-3}$)} &  & {\tiny{}$E_{g}^{dir}$ (eV)} & {\tiny{}$E_{o}$ (eV)} & {\tiny{}$E_{B}$ (eV)} & {\tiny{}$T_{exc}$ (K)}\tabularnewline
\hline 
{\tiny{}Rb$_{2}$TiCl$_{6}$} & {\tiny{}2.74} & {\tiny{}1.56} & {\tiny{}8.36} & {\tiny{}0.17} & {\tiny{}0.17} & {\tiny{}6.65} &  & {\tiny{}5.16} & {\tiny{}4.18} & {\tiny{}0.98} & {\tiny{}11362}\tabularnewline
{\tiny{}Rb$_{2}$SeCl$_{6}$} & {\tiny{}3.42} & {\tiny{}0.54} & {\tiny{}9.18} & {\tiny{}0.07} & {\tiny{}0.39} & {\tiny{}0.53} &  & {\tiny{}4.66} & {\tiny{}4.02} & {\tiny{}0.64} & {\tiny{}7420}\tabularnewline
{\tiny{}Rb$_{2}$PdCl$_{6}$} & {\tiny{}3.29} & {\tiny{}0.68} & {\tiny{}7.04} & {\tiny{}0.15} & {\tiny{}0.32} & {\tiny{}0.97} &  & {\tiny{}3.68} & {\tiny{}3.52} & {\tiny{}0.16} & {\tiny{}1855}\tabularnewline
\hline 
\end{tabular}{\tiny\par}
\end{table}

After determining the $E_{B}$, we also compute several additional
excitonic parameters$-$such as excitonic temperature ($T_{exc}$),
exciton radius ($r_{exc}$), and the probability of a wave function
for the electron-hole pair at zero charge separation\textcolor{red}{{}
}($|\phi_{n}(0)|^{2}$)\citep{chapter2-38,chapter5-16,chapter5-18}.
These parameters are crucial for assessing the efficiency of optoelectronic
devices and are given in Table \ref{tab:4}. The excitonic temperature
($T_{exc}$) is the highest temperature at which an exciton remains
stable. The thermal energy needed to dissociate an exciton is given
by $E_{B}=k_{B}T_{exc}$, where $k_{B}$ is the Boltzmann constant.

The exciton radius ($r_{exc}$) is determined using the following
calculation\citep{chapter2-38,chapter5-16}:

\begin{equation}
r_{exc}=\frac{m_{0}}{\mu^{*}}\mathrm{\varepsilon_{eff}}n^{2}r_{Ry}\label{eq:2}
\end{equation}

Here, $n$ represents the exciton energy level, and $r_{Ry}$ = 0.0529
nm is the Bohr radius. In our study, we consider the electronic contribution
to the dielectric function ($\varepsilon_{\infty}$) as the effective
value, with $n=1$, which yields the smallest exciton radius.

The probability of a wave function at zero charge separation ($|\phi_{n}(0)|^{2}$)
is calculated using the exciton radius as follows\citep{chapter2-38,chapter5-16}:

\begin{equation}
|\phi_{n}(0)|^{2}=\frac{1}{\pi(r_{exc})^{3}n^{3}}\label{eq:3}
\end{equation}

Moreover, the excitonic lifetime ($\tau_{exc}$)\citep{chapter2-38,chapter5-16}
can be qualitatively described as the inverse of $|\phi_{n}(0)|^{2}$.
As a result (see Table \ref{tab:4}), the $\tau_{exc}$ values for
the examined VODPs follow the order: Rb$_{2}$SeCl$_{6}$ $>$ Rb$_{2}$PdCl$_{6}$
$>$ Rb$_{2}$TiCl$_{6}$. A longer exciton lifetime indicates a lower
rate of carrier recombination, which in turn improves the quantum
yield and boosts conversion efficiency. This suggests that the VODPs
with longer exciton lifetimes are well-suited for optoelectronic applications.

\subsection{Polaronic Properties:}

To further deepen our understanding of the fundamental limits of mobility,
it would be advantageous to conduct a first-principles prediction.\textcolor{red}{{}
}To accurately calculate the mobility of polar semiconductors, such
as halide and chalcogenide perovskites, it is crucial to consider
the polaron state. This state arises from the strong interactions
between charge carriers and polar optical phonons, rather than focusing
solely on the free carrier state. This interaction is known as Fr\"ohlich's
interaction and is expressed through the dimensionless Fr\"ohlich parameter
$\alpha$\citep{chapter5-16},

\begin{equation}
\alpha=\frac{1}{4\pi\varepsilon_{0}}\frac{1}{2}\left(\frac{1}{\varepsilon_{\infty}}-\frac{1}{\varepsilon_{static}}\right)\frac{e^{2}}{\hbar\omega_{LO}}\left(\frac{2m^{*}\omega_{LO}}{\hbar}\right)^{1/2},\label{eq:4}
\end{equation}

where, $\varepsilon_{0}$ represents the permittivity of free space,
$\varepsilon_{\infty}$ and $\varepsilon_{static}$ are the electronic
(optical) and static dielectric constants, respectively, $\omega_{LO}$
is the characteristic phonon angular frequency, and $m^{*}$ is the
carrier effective mass. The thermal \textquotedbl B\textquotedbl{}
method described by Hellwarth et al.\citep{chapter2-22} is employed
to determine $\omega_{LO}$, which involves taking the spectral average
across multiple phonon branches. Our results (see Table \ref{tab:5})
indicate that our investigated compounds exhibit intermediate to strong
carrier-phonon coupling ($\alpha$ = 4.02$-$10.05)\citep{chapter2-20}.

It is important to note that polaron formation can lead to a decrease
in the quasiparticle (QP) energies of both electrons and holes. The
polaron energy ($E_{p}$) is estimated by using the value of $\alpha$
as\citep{chapter5-16,chapter5-18}:

\begin{equation}
E_{p}=(-\alpha-0.0123\alpha^{2})\hbar\omega_{LO}\label{eq:5}
\end{equation}

For Rb$_{2}$TiCl$_{6}$, Rb$_{2}$SeCl$_{6}$, and Rb$_{2}$PdCl$_{6}$,
the QP gap is lowered by 0.49, 0.28, and 0.23 eV, respectively. On
comparing these values with $E_{B}$ (see Table \ref{tab:4}), the
charge-separated polaronic states of Rb$_{2}$TiCl$_{6}$ and Rb$_{2}$SeCl$_{6}$
are found to be less than the bound exciton, whereas the opposite
trend is observed for Rb$_{2}$PdCl$_{6}$ VODP.

Other polaron parameters, such as the effective mass of the polaron
and polaron mobility, are also essential for optoelectronic applications.
Feynman's extended version of Fr\"ohlich's polaron theory (for small
$\alpha$) is employed to derive the effective mass of the polaron
($m_{p}$) as follows:\citep{chapter2-23}:

\begin{equation}
m_{p}=m^{*}\left(1+\frac{\alpha}{6}+\frac{\alpha^{2}}{40}+...\right),\label{eq:6}
\end{equation}

where, $m^{*}$ represents the effective mass of charge carrier determined
through band structure calculations. It is observed that the polaronic
masses of the investigated VODPs are 2$\lyxmathsym{\textendash}$4
(4$\lyxmathsym{\textendash}$5) times greater than the effective masses
of electrons (holes). This confirms the enhanced carrier-lattice interactions,
which in turn explains the relatively lower charge carrier mobility
compared to nonpolar or less polar perovskites.

Using the Hellwarth polaron model\citep{chapter2-22}, the polaron
mobility ($\mu_{p}$) is expressed as follows:

\begin{equation}
\mu_{p}=\frac{\left(3\sqrt{\pi}e\right)}{2\pi c\omega_{LO}m^{*}\alpha}\frac{\sinh(\beta/2)}{\beta^{5/2}}\frac{w^{3}}{v^{3}}\frac{1}{K(a,b)}\label{eq:7}
\end{equation}

where, $e$ is the charge of electron, $\beta=hc\omega_{LO}/k_{B}T$,
and $w$ and $v$ are the temperature-dependent variational parameters,
and $K(a,b)$ is a function of $\beta$, $w$, and $v$, and defined
as follows:
\begin{center}
\begin{equation}
K(a,b)=\int_{0}^{\infty}du\left[u^{2}+a^{2}-b\cos(vu)\right]^{-3/2}\cos(u)
\end{equation}
\par\end{center}

Here, $a^{2}$ and $b$ are evaluated as:
\begin{center}
\begin{equation}
a^{2}=(\beta/2)^{2}+\frac{(v^{2}-w^{2})}{w^{2}v}\beta\coth(\beta v/2)
\end{equation}
\par\end{center}

\begin{center}
\begin{equation}
b=\frac{v^{2}-w^{2}}{w^{2}v}\frac{\beta}{\sinh(\beta v/2)}
\end{equation}
\par\end{center}

Table \ref{tab:5} shows that Rb$_{2}$SeCl$_{6}$ and Rb$_{2}$PdCl$_{6}$
perovskites exhibit higher polaron mobility for electrons at 300 K
compared to Rb$_{2}$TiCl$_{6}$, which can be attributed to their
smaller carrier effective mass and higher electronic dielectric constant.
On the other hand, all the VODPs show lower polaron mobility for holes
compared to electrons, suggesting that these Rb$_{2}$BX$_{6}$ compounds
behave as n-type semiconductors.

\begin{table}[H]
\caption{\label{tab:5}Polaron parameters corresponding to electrons (e) and
holes (h) of Rb$_{2}$BCl$_{6}$ (B = Ti, Se, Pd) VODPs. Here, $\omega_{LO}$
represents the characteristic phonon angular frequency, $\alpha$
represents Fr\"ohlich interaction parameter, $E_{p}$ represents the
polaron energy, $m_{p}$ represents effective mass of the polaron
(in terms of $m^{\ast}$), and $\mu_{p}$ represents the polaron mobility.}

\centering{}{\small{}}%
\begin{tabular}{ccccccccccccc}
\hline 
\multirow{2}{*}{{\small{}Compounds}} & \multirow{2}{*}{{\small{}$\omega_{LO}$ (THz)}} & \multicolumn{2}{c}{{\small{}$\alpha$}} &  & \multicolumn{2}{c}{{\small{}$E_{p}$ (eV)}} &  & \multicolumn{2}{c}{{\small{}$m_{p}/m^{*}$}} &  & \multicolumn{2}{c}{{\small{}$\mu_{p}$ (cm$^{2}$V$^{-1}$s$^{-1}$)}}\tabularnewline
\cline{3-4} \cline{4-4} \cline{6-7} \cline{7-7} \cline{9-10} \cline{10-10} \cline{12-13} \cline{13-13} 
 &  & {\small{}$e$} & {\small{}$h$} &  & {\small{}$e$} & {\small{}$h$} &  & {\small{}$e$} & {\small{}$h$} &  & {\small{}$e$} & {\small{}$h$}\tabularnewline
\hline 
{\small{}Rb$_{2}$TiCl$_{6}$} & {\small{}5.76} & {\small{}8.30} & {\small{}10.05} &  & {\small{}0.22} & {\small{}0.27} &  & {\small{}4.11} & {\small{}5.20} &  & {\small{}0.31} & {\small{}0.11}\tabularnewline
{\small{}Rb$_{2}$SeCl$_{6}$} & {\small{}4.88} & {\small{}4.14} & {\small{}8.47} &  & {\small{}0.09} & {\small{}0.19} &  & {\small{}2.12} & {\small{}4.21} &  & {\small{}5.88} & {\small{}0.22}\tabularnewline
{\small{}Rb$_{2}$PdCl$_{6}$} & {\small{}4.53} & {\small{}4.02} & {\small{}7.52} &  & {\small{}0.08} & {\small{}0.15} &  & {\small{}2.07} & {\small{}3.66} &  & {\small{}5.78} & {\small{}0.37}\tabularnewline
\hline 
\end{tabular}{\small\par}
\end{table}

\section{Conclusions:}

In this paper, we present a comprehensive study to investigate the
effects of B-site atoms on the structural, mechanical, electronic,
optical, excitonic, and polaronic properties of Rb$_{2}$BCl$_{6}$
(B = Ti, Se, Ru, Pd) VODPs using state-of-the-art ground- and excited-state
methods. Our study reveals that these compounds maintain the standard
face-centered cubic lattice with the space group Fm$\bar{3}$m (No.
225). The stability of these perovskites is confirmed by their phonon
band structures and elastic properties. The GW-based electronic structure
calculations predict that Rb$_{2}$BCl$_{6}$ (B = Ti, Se, Pd) VODPs
exhibit a wide range of bandgaps, varying from 3.63 to 5.14 eV, whereas
Rb$_{2}$RuCl$_{6}$ displays a metallic character. Subsequently,
by solving the Bethe-Salpeter equation (BSE), the optical properties
are accurately estimated and these VODPs are shown to exhibit absorption
in the near-ultraviolet to mid-ultraviolet spectrum. These materials
also possess low to high exciton binding energies (0.16-0.98 eV) and
they have a negligible ionic contribution to the effective dielectric
screening used to calculate the excitonic parameters. In addition,
Fr\"ohlich\textquoteright s mesoscopic model reveals that intermediate
to strong carrier-phonon coupling is present and cannot be overlooked
in these materials, leading to reduced charge carrier mobility. Overall,
the fundamental understanding gained from this study not only highlights
that Rb$_{2}$BCl$_{6}$ VODPs are a promising class of stable and
nontoxic materials for next-generation optoelectronic devices, but
also offers valuable insights into the exploration and prediction
of future light-harvesting materials.
\begin{acknowledgments}
S.A. would like to acknowledge the Council of Scientific and Industrial
Research (CSIR), Government of India {[}Grant No. 09/1128(11453)/2021-EMR-I{]}
for financial support. The high performance computing facility \textquotedbl Magus\textquotedbl{}
and workstations available at the School of Natural Sciences, Shiv
Nadar Institution of Eminence (SNIoE), were used to perform all calculations.
\end{acknowledgments}

\bibliographystyle{apsrev4-2}
\bibliography{refs}

\end{document}